\documentclass[12pt]{iopart}

\usepackage{circuitikz} 
\usepackage{iopams}
\usepackage{cite}

\begin{document}

\title[]{Phase slips dynamics in gated Ti and V all-metallic supercurrent nano-transistors: a review.}

\author{C. Puglia$^{1,2,*}$, G. De Simoni$^{2}$, F. Giazotto$^{2}$}

\address{Dipartimento di Fisica, Università di Pisa, Largo Bruno Pontecorvo 3, I-56127 Pisa, Italy\\$^2$NEST, Istituto Nanoscienze-CNR and Scuola Normale Superiore, I-56127 Pisa, Italy}
\ead{$^*$claudio.puglia@df.unipi.it}
\vspace{10pt}
\begin{indented}
\item[]February 2021
\end{indented}

\begin{abstract}
The effect of electrostatic gating on metallic elemental superconductors was recently demonstrated in terms of modulation of the switching current and control of the current phase relation in superconducting quantum interferometers. The latter suggests the existence of a direct connection between the macroscopic quantum phase ($\phi$) in a superconductor and the applied gate voltage. The measurement of the switching current cumulative probability distributions (SCCPD) is a convenient and powerful tool to analyze such relation. In particular, the comparison between the conventional Kurkijärvi–Fulton–Dunkleberger model and the gate-driven distributions give useful insights into the microscopic origin of the gating effect. In this review, we summarize the main results obtained in the analysis of the phase slip events in elemental gated superconducting weak-links in a wide range of temperatures between 20 mK and 3.5 K. Such a large temperature range demonstrates both that the gating effect is robust as the temperature increases, and that fluctuations induced by the electric field are not negligible in a wide temperature range. 
\end{abstract}

%\maketitle

\section{Introduction}
Although static electric fields are largely ineffective on metals due to their large free carrier density \cite{Lang1970,Larkin1963,Ummarino2017,Shapiro1984,Burlachkov1993,Lee1996,Morawetz2008,Lipavsky2006}, it has been unexpectedly shown that the superconducting properties of metallic Bardeen-Cooper-Schrieffer (BCS) superconducting wires \cite{DeSimoni2018}, Dayem bridges \cite{Paolucci2018,Paolucci2019}, and proximity superconductor-normal metal-superconductor (SNS) \cite{DeSimoni2019} Josephson junctions (JJ) can be manipulated via the application of a control gate voltage. The critical current ($I_C$) of these systems was observed to decrease as a function of an increasing control gate voltage, while no clear relation could be established with an eventual current flowing between the gate and the superconductor \cite{Rocci2020}. The gate voltage was found as well to affect the current-phase relation of superconducting quantum interference devices (SQUIDs) \cite{Paolucci2019b}, on the one hand, through direct suppression of the critical current of a gated weak-link. On the other hand, phase shifts in the SQUID current vs flux relation were observed also for gate voltages so small that no apparent influence on $I_C$ was observed. Such behavior was interpreted as stemming from the voltage-driven occurrence of fluctuations of the superconducting phase ($\phi$) \cite{Paolucci2019b}. This hypothesis has found qualitative support through the study of the evolution of the dynamics of the phase slips, i.e., local random 2$\pi$ phase jumps that responsible for the superconducting-to-normal state switching \cite{Pekker2009,Foltyn2015}. The investigation of the switching current ($I_S$) probability distribution (SCPD) of a JJ is, indeed, a tool to access information on the phase slippage dynamics as a function of external parameters such as the temperature or externally applied fields \cite{Bezryadin2000,Bezryadin2010,Bezryadin2012,Zgirski2018}, and allowed us to demonstrate that in metallic superconductors under the action of a gate voltage the SCPD is remarkably different from that measured as a function of the temperature. This suggests, thereby, that the gate-voltage driven suppression of $I_C$ cannot be related to a conventional thermal-like overheating of the superconductor, but rather it involves a different elusive mechanism which is yet to be identified, and fully understood \cite{Virtanen2018,Solinas2020,Mercaldo2020}.

In this manuscript, we examine in-depth the modification of the SCPD of gated all-metallic Dayem nano-bridge devices for temperatures as high as 3.75 K, i.e., in a temperature range much wider than that explored in previous analogous experiments. To this aim, we report on Ti and V gated all-metallic supercurrent nano-transistors and discuss the evolution of the switching escape rate in the presence of a control gate voltage. Our data provide another tile in the puzzle of the interaction of electric fields and the BCS condensate, and are particularly relevant, for instance, in view of the realization of electrostatically-driven phase-slip qubit based on gated metallic Josephson nanojunctions \cite{Mooij2005}.

\section{RCSJ model for phase slips in superconducting weak-links}

The resistively capacitively shunt junction (RCSJ) model \cite{Barone1982} is a convenient tool to analytically describe phase-slips phenomena in a JJ. It describes the behavior of a JJ under an arbitrary current bias. In this scheme, the Josephson weak-link is replaced by the parallel of a resistor $R$, a capacitor $C$ and a non-linear current generator $I(\phi)$. For systems with a sinusoidal current-phase relation (CPR), $I(\phi,T)=I_S(T)\sin(\phi)$ \cite{Josephson1962}.

\begin{figure}[!ht]
\centering
\includegraphics[width=\textwidth]{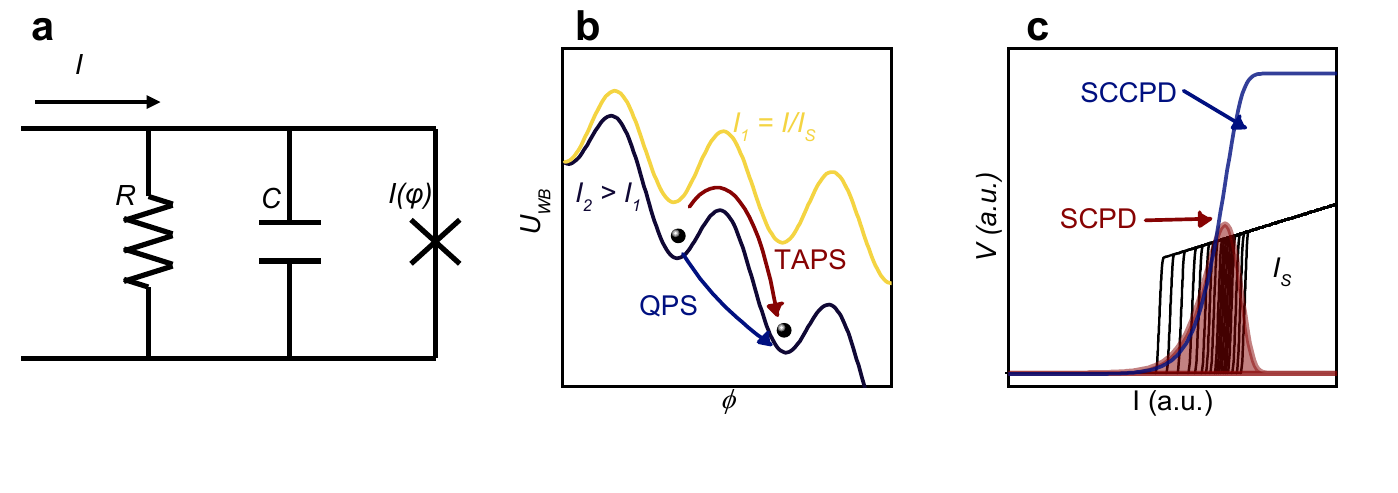}
\caption{a. RCSJ model in which the junction is represented as the parallel of a resistor, a capacitor and a phase-dependent current generator. b. Particle in a tilted washboard potential. The "jump" between two adjacent minima can occur through two different mechanisms, either thermal activated hopping over the barrier (red) or quantum tunnelling (blue). The ratio $\frac{I}{I_S}$ determines the potential tilting. c. Repeated $I$ vs $V$ measurements. The areas most populated with transitions are the regions where it is more likely for the bias current to trigger a transition to the normal state. The red curve represents the probability of transition for bias current $I$, the SCPD. The blue curve is obtained via integration of the SCPD and is the cumulative probability distribution, the SCCPD.}
\label{fig:RCSJ}
\end{figure}

%\begin{figure}[!ht]
%\centering
%\includegraphics[width=\textwidth]{images/washboard.pdf}
%\caption[Tilted washboard potential]{Particle in a tilted washboard potential. The "jump" between two %adjacent minima can occur with two different mechanism, thermal activate hopping of the barrier (red) %and quantum tunneling (blue). The ratio $\frac{I}{I_S}$ determines the potential tilting.}
%\label{fig:wash}
%\end{figure}

This phenomenological circuital approach allows to write down the differential equation
 $$ I=I_S \sin{\left(\phi\right)}+\frac{\phi_0}{2\pi R_N}\dot\phi+C\frac{\phi_0}{2\pi }\ddot\phi, $$
 where $R_N$ is the normal-state resistance of the junction, $C$ its capacitance, $\phi_0$ is the magnetic flux quantum, and  $I$ is the circulating current. Such equation is equivalent to the motion of a classical particle of mass $m_p=\frac{C\hslash^2}{4e^2}$ under the action of a tilted washboard potential $U_{WB}$, as depicted in Fig. \ref{fig:RCSJ}b \cite{Barone1982}. $U_{WB}$ is characterized by a series of minima separated by barriers with a height expressed by \cite{Tinkham2004}
 $$ \Delta U(I,T)= a\ E_J(T)\left(1-\frac{I}{I_S(T)}\right)^b,$$
where $a$ and $b$ take into account the geometry of the systems.

The motion of a phase particle from one local minimum to the successive one leads to sharp 2$\pi$ variations of $\phi$. Such events are called phase slips (PSs) and are attributed to two distinct mechanisms \cite{Bezryadin2010}: thermal hopping \cite{Longobardi2011} and quantum tunnelling. The former requires a thermal activation energy, while the latter is only weakly dependent on temperature. The interplay of the two contributions allows us to define three different regimes divided by two crossover temperatures $T_Q<T_M$:
\begin{itemize}
  \item for $T<T_Q$ the quantum tunnelling provides the majority of the phase slip events and the system lays in the quantum phase slip (QPS) regime \cite{Zaikin1997,Sahu2009};
  \item for intermediate temperatures, when $T_Q<T<T_M$, the thermal energy is large enough to trigger the hopping of the barrier of the washboard potential, leading the system in the thermally-activated phase slip (TAPS) regime;
  \item for higher temperatures, $T>T_M$, the phase slips events occur more than one at once, and the system shows the multiple phase slip (MPS) regime \cite{Ejrnaes2019}. In contrast to QPS and TAPS, there are no analytical models for MPS.
\end{itemize}

For a particle in an equilibrium position, we can define an escape rate $\Gamma$, which evaluates the survival time of the particle in the equilibrium point. Kramers' theory \cite{Kramers1940} provides a general formula for the escape rate in TAPS \cite{Zaikin1997,Golubov1999} and QPS regimes \cite{Giordano1988,Bezryadin2000}
$$\Gamma_{TAPS}(I,T)=\frac{L}{2\pi \xi(T)\tau_{GL}(T)}\sqrt{\frac{\Delta U(I,T)}{k_B T}}\exp{\left(-\frac{\Delta U(I,T)}{k_B T}\right)}$$
$$\Gamma_{QPS}(I,T,T_{QPS})=\frac{L}{2\pi \xi(T)\tau_{GL}(T)}\sqrt{\frac{\Delta U(I,T)}{k_B T_{QPS}}}\exp{\left(-\frac{\Delta U(I,T)}{k_B T_{QPS}}\right)},$$
respectively, where $L$ is the length of the junction, $\xi(T)$ is the coherence length, $\tau_{GL}=\frac{\pi \hslash}{2k_B (T_C-T)}$ is the Ginzburg-Landau time constant \cite{Bezryadin2012} and $T_{QPS}$ is effective temperature that allows to compare the height of the barrier $\Delta U$ with the energy term $k_B T_{QPS}$.
Since $\Delta U$ is a function of the weak-link bias current, the switching probability due to the occurrence of a phase slip event as a function of $I$ can be reconstructed by acquiring a large number of $V$ vs $I$ characteristics. This procedure shows that the transition probability is distributed with an asymmetric bell-shaped function around a value which is conventionally defined the as the critical current $I_C$ of the weak-link.

%\begin{figure}[!ht]
%\centering
%\includegraphics[width=\textwidth]{images/multipleivs.pdf}
%\caption[Repeated $I$ vs $V$ curves measures]{Repeated $I$ vs $V$ measures. The most populated areas are the region where is more likely for the bias current to trigger a transition to the normal state.}
%\label{fig:iv}
%\end{figure}

The connection between the probability distribution $P(I)$ and the phase slip rate $\Gamma$ is provided by the Kurkijärvi theory (KT) \cite{Bezryadin2012,Kurkijarvi1972,Fulton1971,Garg1995} via the  Kurkijärvi–Fulton–Dunkleberger (KFD) transformation, which for continuous and discrete distributions can be written as 
$$\Gamma(I)=P(I)\nu_I\left[1-\int_0^IP(I')dI'\right]^{-1}\ \ \ \ \ \ \ \  \Gamma (I_N,T)=\frac{P(I_N,T) \nu_I}{1-w\sum_{k=0}^N P(I_k,T)},$$
where $\nu_I$ is the slope of the current ramp, $w$ is the bin size of the $P(I,T)$ histogram, and $P(I_k,T)$ is the switching probability in the current interval $[k w, (k+1)w]$ with $k\in {N}$.

It is worth to notice that the different mechanism for the escaping phenomena leads to a non-trivial behavior of the distribution width. In particular, below $T_{Q}$ where the tunnelling effects are predominant, the standard deviation $\sigma$ vs $T$ is expected to be temperature independent because QPSs do not require a thermal activation. On the other hand, when $T_Q<T<T_M$,  $\sigma$ is expected to grow as a function of $T$ because of the larger available thermal energy that allows hopping events for a wider range of bias current values. Finally, for $T>T_M$, the standard deviation is expected to decrease as a function of the temperature, as already observed in similar experiments \cite{Bezryadin2010,Sahu2009,Longobardi2011}.

Such a model, although phenomenological, is very effective and allows to analyze the effect of the temperature in the dynamics of the phase slip events in mesoscopic Josephson junctions. Moreover, it is suitable to study how such dynamics is modified when the weak link is subject to the action of electrostatic gating. In the following, we will focus on this specific issue by reporting the results of experiments in which the modifications of the phase slips occurrence in nanojunctions was obtained via the application of a control gate voltage. Data are then discussed and interpreted in the framework of the conventional theory. 

\section{Titanium Dayem bridge PS dynamics}

\begin{figure}[!ht]
\centering
\includegraphics[width=\textwidth]{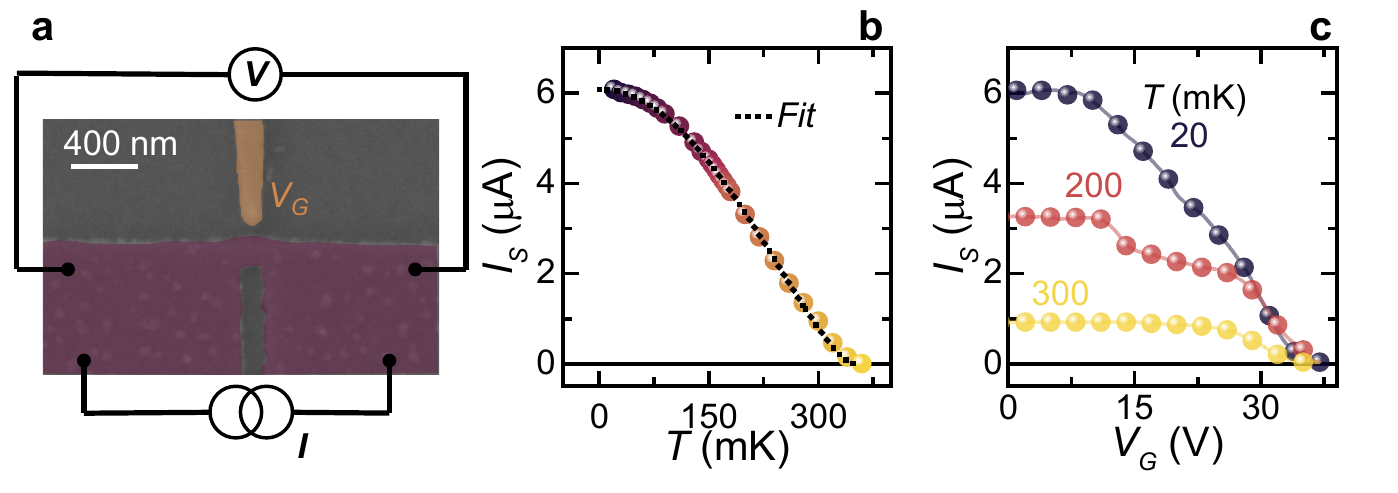}
\caption{a. Pseudo-color SEM image of a representative titanium gated (orange) Dayem bridge (purple) with the four-probes bias scheme used for the characterization. b. Critical current $I_S$ vs $T$ characteristic fitted with the conventional Bardeen's formula \cite{Bardeen1957} (black dotted line). The fit parameters are $T_C \simeq 348$ mK and $I_S^0 \simeq 6.02\ \mu$A. c. $I_S$ vs $V_G$ characteristics at different bath temperatures between 20 mK and 300 mK. Switching current values were computed averaging over 50 acquisitions.}
\label{fig:tidev}
\end{figure}

Ti wires and Dayem bridges transistors are the systems in which the gating effect on metallic superconductors was originally demonstrated \cite{DeSimoni2018}. In this section, we discuss the effect of electrostatic gating on these systems in terms of its impact on the switching current probability distribution.  Data shown in the following were acquired on a 30-nm-thick ultra high-vacuum e-beam evaporated titanium Dayem bridge (150-nm-long and 120-nm-wide) shaped with a single-step electron beam lithography of a poly-methyl-methacrylate (PMMA) polymeric mask deposited on top of a sapphire (Al$_2$O$_3$) substrate. In the same evaporation, a 140-nm-wide gate electrode separated by a distance of approximately 80 nm from the constriction was deposited. A pseudo-color scanning electron micrograph (SEM) of a typical device is shown in Figure~\ref{fig:tidev}a.

The device shows a normal-state resistance of 550 $\Omega$ and a switching current of 6 $\mu$A at a temperature of 20 mK. The evolution of the switching current as a function of bath temperature is shown in Figure \ref{fig:tidev}b. $I_S$ evolution follows the Bardeen's law \cite{Bardeen1962} (dotted line in Figure \ref{fig:tidev}b), $I_S\left(T\right)  =  I_S^0\left[1-\left(T/T_C\right)^2\right]^{3/2}$, where $I_S^0$ is the zero-temperature switching current, and $T_C$ is the Ti film critical temperature. The fit procedure returns as parameters $ T_C \simeq 348 $ mK and $ I_S^0 \simeq 6.02\ \mu$A that are in agreement with similar devices values \cite{Paolucci2018}. Similarly, the evolution of the switching current as a function of the gate voltage applied through the gate electrode is shown in Figure~\ref{fig:tidev}c. The curves, acquired at different bath temperatures ranging from 20 mK to 300 mK, show monotonic suppression of $I_S$, with complete quenching for $ V^C_G \simeq 34 $ V. Note the conventional widening of the plateau for which $I_S$ is unaffected by the gate voltage as the temperature increases, in agreement with experiments performed on gated elemental superconductors \cite{Paolucci2018,DeSimoni2018,DeSimoni2019,Paolucci2019,Bours2020}.

\begin{figure}[!ht]
\centering
\includegraphics[width=\textwidth]{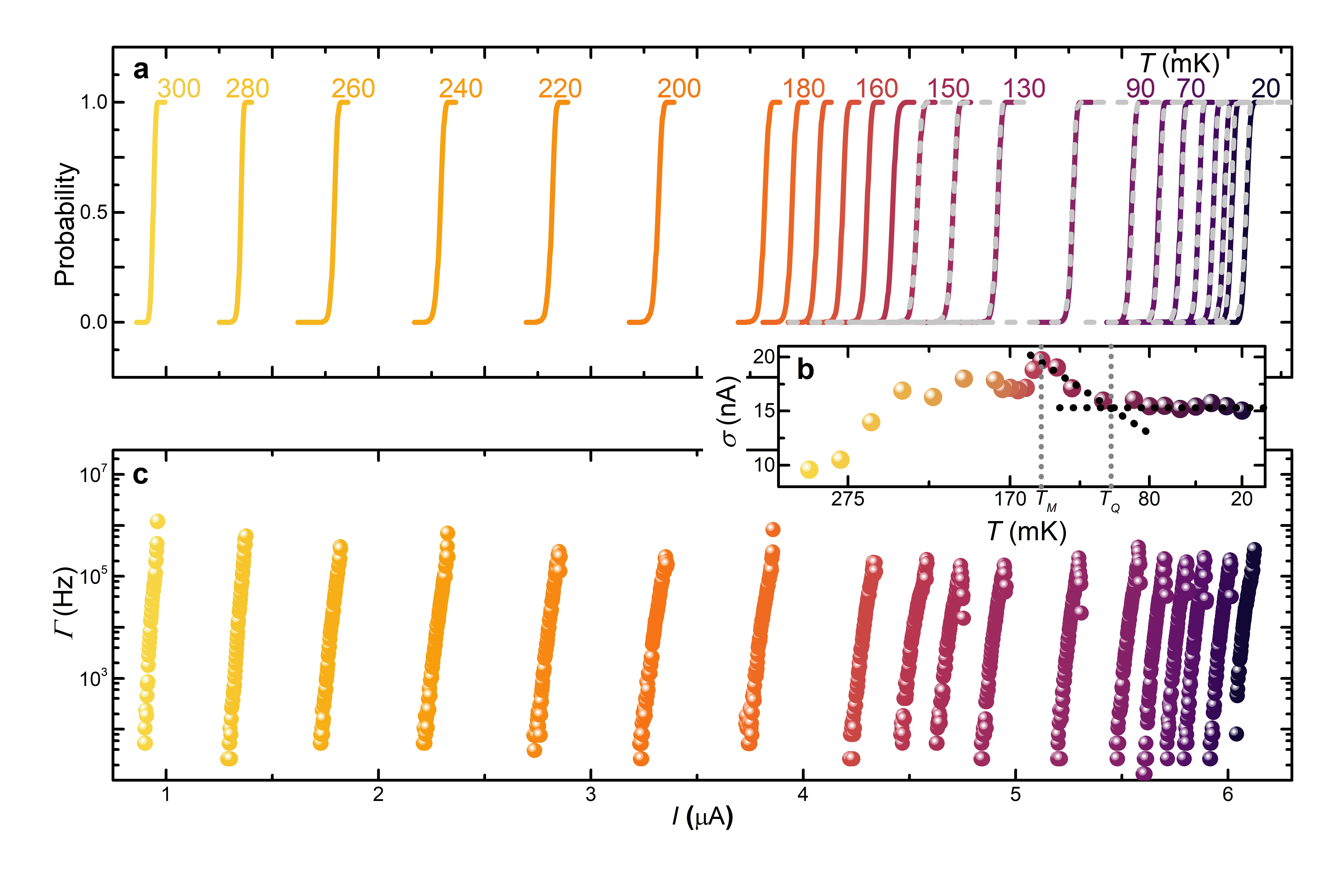}
\caption{a. SCCPDs computed at several bath temperature ranging from 20 mK to 300 mK. $I_S$ was acquired $10^5$ times to reconstruct the SCPD. Dashed grey lines represent the fit curves obtained for QPS and TAPS regimes in the framework of the KFD theory \cite{Puglia2020}. b. $\sigma$ vs $T$ characteristic with the two crossover temperatures $T_Q\simeq110$ mK and $T_M\simeq 150$ mK that define the transition between QPS/TAPS and TAPS/MPS regimes, respectively. c. Escape rate computed from the data in panel a with the KFD transform. They show a phase lifetime between 1 $\mu$s and 10 ms.}
\label{fig:tithermal}
\end{figure}

The effect of the temperature on the number of phase slip events was assessed by acquiring $10^5$ switching current values biasing the junction with a linear current ramp in the four-probe scheme shown in Figure \ref{fig:tidev}a at different temperatures ranging from 20 mK to 300 mK. $I_S$ values were acquired with a  750 KHz bandwidth input/output analogue-to-digital/digital-to-analogue converter (ADC/DAC) board for the acquisition of the voltage drop across the junction and the generation of the bias current, respectively. The current input signal consisted of an 8.7 Hz sawtooth wave obtained by applying a  voltage signal generated by the digital board to a 1 M$\Omega$ load resistor.  The resulting signal consisted of a positive linear ramp with amplitude $10\ \mu$A, and slope $\nu_I = 133\ \mu$A/s followed by a 100 ms zero-current plateau essential for the system to cool down between two consecutive super-to-normal state transitions. $I_S$ acquisitions were combined via numerical integration to reconstruct the switching current cumulative probability distributions (SCCPDs, the S-curves), as shown in Figure \ref{fig:tithermal}a. These curves represent, for a given current $I$, the probability that the switching current satisfies the condition $I_S<I$.

We note that, as conventionally observed in similar systems \cite{Bezryadin2010}, the S-curves shift to a lower value of injected current as the temperature $T$ increases. Such observation is equivalent to the decrease of the critical temperature showed in Figure~\ref{fig:tidev}b. The width of the SCCPDs ($\sigma$) can be estimated by computing the standard deviation of the switching current probability distribution (SCPD) that is obtained via numerical derivation of the SCCPDs. The conventional behavior of the $\sigma$ vs $T$ characteristics is assessed in Figure~\ref{fig:tithermal}b where the three different phase slip regimes can be distinguished,  separated by the crossover temperatures $ T_Q \simeq 110 $ mK and $ T_M \simeq 150 $ mK: 
\begin{itemize}
    \item QPS regime: for $T<T_Q$, the standard deviation is constant as a function of the temperature since the quantum tunnelling process does not require activation energy.
    \item TAPS regime: for $T_Q<T<T_M$, $\sigma$ and $T$ are linearly correlated because the temperature increase provides a growing amount of thermal energy to the systems, facilitating the hopping of the potential barrier. 
    \item MPS regime: for $T>T_M$, the width of the transition decreases, as already observed in previous systems.
\end{itemize}
 The escape rate $\Gamma(I,T)$ is another relevant quantity to be extracted from SCCPDs. It provides an estimate of the phase lifetime in the Josephson nanojunctions \cite{Fulton1974,Bae2012}, and ranges between $1\ \mu$s ($\Gamma \sim10^6 $ Hz) and 10 ms ($\Gamma\sim10^2$ Hz). The values obtained for the escape rate are in agreement with the literature data \cite{Bezryadin2010,Bezryadin2012,Blackburn2016} performed on superconducting weak-links.

\begin{figure}[!ht]
\centering
\includegraphics[width=\textwidth]{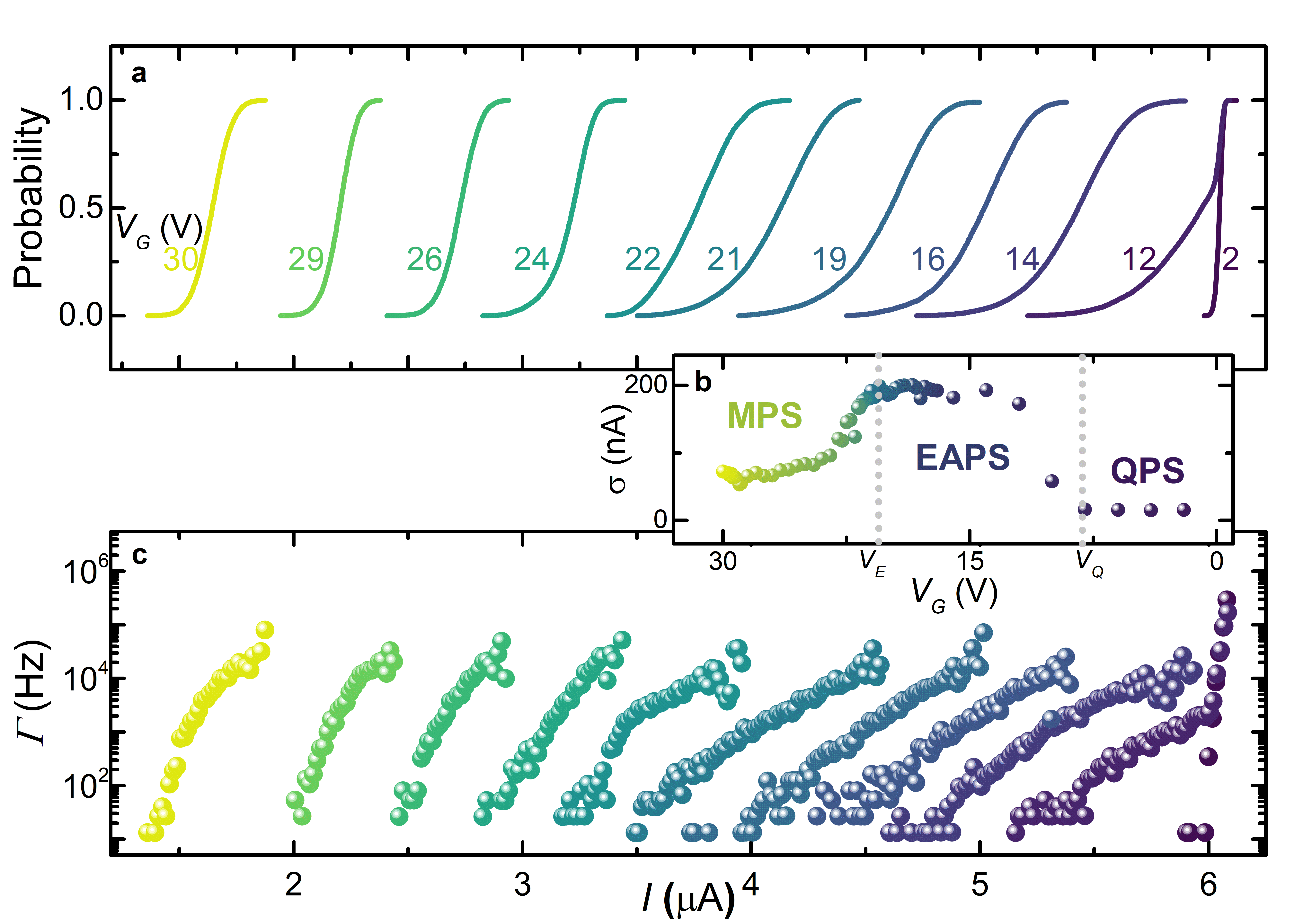}
\caption{a. SCCPDs acquired at several gate voltages ranging from 2 V to 30 V.  b. $\sigma$ vs $T$ characteristic with the two crossover voltages $V_E$ and $V_Q$ that define the transition between QPS/EAPS and EAPS/MPS regimes, respectively. c. Escape  rates  computed  from  the  data  in  panel  a.   They  shows  a  phase  lifetime between 10 $\mu$s and 100 ms.}
\label{fig:tielectric}
\end{figure}

The effect of electrostatic gating on the shape of SCCPDs is shown in Figure~\ref{fig:tielectric}a. The curves were acquired at 20 mK for several gate voltage values ranging from 2 V to 30 V. The shape of the S-curves is substantially modified by gating. Indeed, for $ V_G < 8 $ V the curves are almost identical to the zero-gate one, but a non-zero probability transition for low current values appears for $ 8$ V$< V_G < 14 $ V. In the range between $14$ V $< V_G < 24$ V the curves widens significantly due to the increase of the phase slip events. Finally, at higher gate voltages ($V_G > 24$ V), the S-curve narrows.

The evolution of the shape of SCCPDs is plotted in Figure~\ref{fig:tielectric}b with the $\sigma$ vs $V_G$ characteristic. Notably, a region of constant standard deviation is observed, stating a  minor contribution of gate effect to the phase slips occurrence for low $V_G$ values. In analogy with the thermal case, such evolution seems to be equivalent to the QPS regime.  By increasing $V_G$ the standard deviation grows up to a value of about 200 nA. This region is defined as an electrically-activated phase slip (EAPS) regime due to the increase of gate-induced phase slips. At higher gate voltage, $\sigma$ decreases and saturates to approximately 75 nA. Despite resembling the MPS regime, such behavior cannot be ascribed to conventional heating effects. The evolution of $\sigma$ vs $V_G$ defines two crossover gate voltages $ V_Q \simeq 8 $ V and $ V_E \simeq 21 $ V that represent the transition between QPS/EAPS and EAPS/MPS regimes, respectively.

It is worth to emphasize that the lifetime of the phase particle, represented in Figure~\ref{fig:tielectric}c as $\Gamma$, in the gate-driven regimes ranges between $10 \ \mu$s and 100 ms and on average is around one order of magnitude smaller than the thermal rates. This indicates a possible action of the electrostatic gating in driving the superconductor far from equilibrium. 

\begin{figure}[!ht]
\centering
\includegraphics[width=\textwidth]{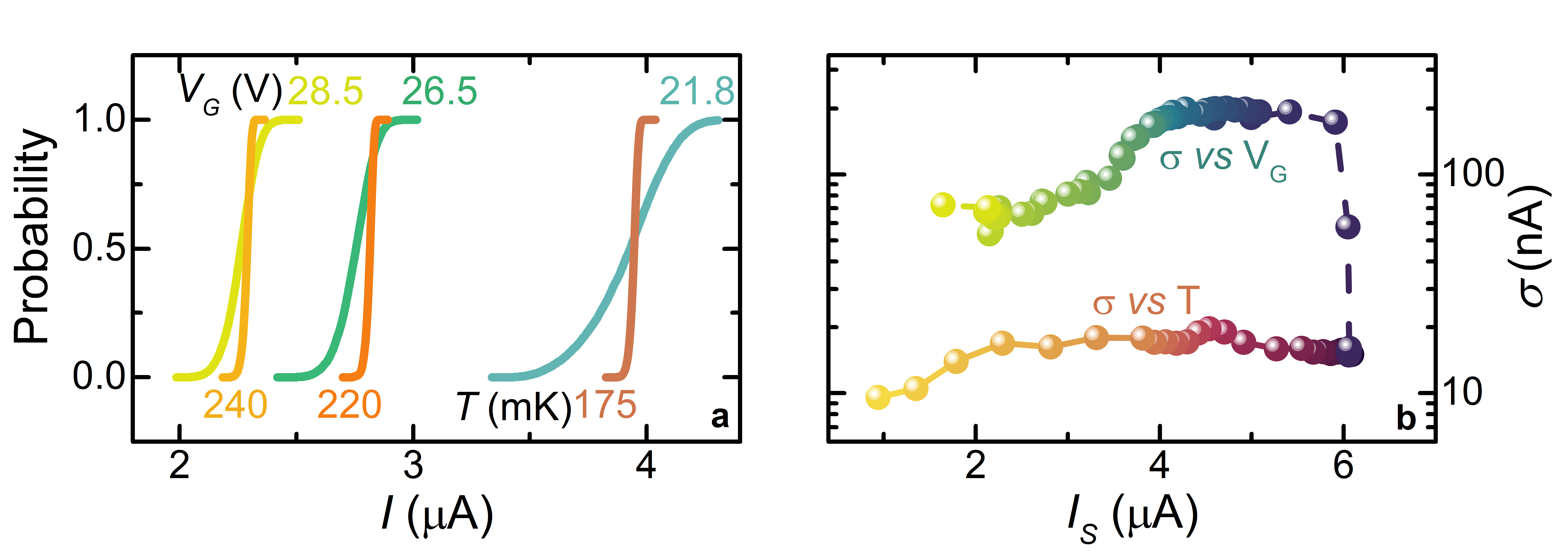}
\caption{a. $\sigma$ vs $I_C$ curves as a function of temperature at  $V_G=0$ (violet to orange), and vs gate voltage at 20 mK (blue to yellow). b. $I_C$-matched S-curves, brown and orange distributions were acquired for $ V_G = 0 $ at selected temperatures whereas  blue and green characteristics were measured at $ T = 20 $ mK for different gate voltage values.}
\label{fig:ticomparison}
\end{figure}

The differences between thermal and electric S-curves are highlighted by comparing the characteristics with the same critical current $2.2,\ 2.8,\ 4.0\ \mu$A acquired at $V_G=0$ V (brown to orange) and $T=20$ mK (blue to yellow). Figure~\ref{fig:ticomparison}a highlight the differences in the shape of the distributions and in particular the larger width of the gate-induced ones. Such a behavior is probably due to a gate-driven strong non-equilibrium state induced in the superconducting junction. Also, the direct comparison between the $\sigma(I_S(V_G))$ and the $\sigma(I_S(T))$characteristics (reported in Figure~\ref{fig:ticomparison}b) displays a value of the standard deviation of the gate-driven curve larger by one order of magnitude in the first case. We emphasize that such a strong deviation from the thermal case prevents to exploit the conventional theory \cite{Bezryadin2012} to extract an effective electronic temperature in the weak link, which should be much higher than the Ti critical temperature. A result, therefore, completely devoid of physical meaning. This observation suggests that the gate action affects deeply the system phase dynamics resulting in an increased switching probability in a wider bias current range with respect to the corresponding thermal case. This behavior is compatible with the picture of gate-induced phase fluctuations in the Dayem bridge \cite{Paolucci2019b}, and with the formation of a glassy phase configuration along the supercurrent path \cite{Mercaldo2020}.

%The reported experiment analyzes the occurrence of different phase slips regimes in a titanium Dayem bridge Josephson junction in a low-temperature regime down to 20 mK. First of all, the system behaves accordingly with the conventional phase slip theory as a function of the bath temperature. Secondly, the shape of the S-curves is dramatically affected by the applied gate voltage. More specifically, the transition between a zero to unitary probability state occurs in a wider current range. The strong gate-induces non-equilibrium state is charged for the differences observed between thermal and electric CSCPD. Finally, the impossibility to describe the effect of the conventional gating with an effective temperature in the standard theory demonstrate that a thermal origin of the suppression of the supercurrent in extremely unlikely.

\section{Vanadium Dayem bridge PS dynamics}

The experiment performed on Ti weak-links allowed us to study the phase slips dynamics in a temperature range $ T < 300 $ mK where thermal fluctuations and the electron-phonon coupling in the superconductor are strongly suppressed. To assess the influence of the gate-induced phase slip events in a different temperature range the same analysis has to be performed on devices based on different elemental superconductors with a higher critical temperature. This is the case of gate-controllable vanadium (V) Dayem bridges, in which the higher $T_C$ allowed to probe the SCCPD up to bath temperatures above 2 K.

\begin{figure}[!ht]
\centering
\includegraphics[width=\textwidth]{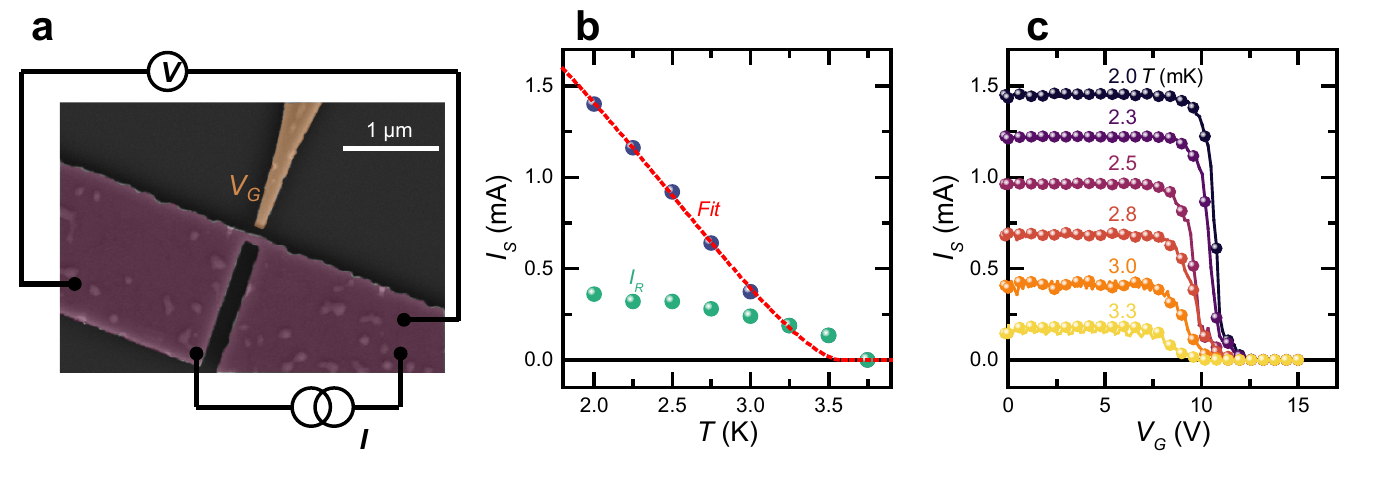}
\caption{a. Pseudo-color SEM image of a representative vanadium gated (orange) Dayem bridge (purple) with the four-probes bias scheme used for the low temperature characterization. b. $I_S$ vs $T$ characteristic fitted with the conventional Bardeen's formula \cite{Bardeen1957} (red dotted line). c. $I_S$ vs $V_G$ characteristics measured at different bath temperatures ranging  between 2 K and 3.3 K. The switching current values were computed by averaging over 50 acquisitions.}
\label{fig:vandev}
\end{figure}

The vanadium gated device, shown in Figure~\ref{fig:vandev}a, consists of a 60-nm-tick, 160-nm-long, 90-nm-wide constriction aligned with a 70-nm-far, and 120-nm-wide side-gate. The metal deposition was performed on a silicon/silicon-dioxide (Si/SiO$_2$) substrate at a rate of 0.36 nm/s in an ultra-high vacuum e-beam evaporator with a base pressure of $\sim 10^{-11}$ Torr \cite{Garcia2009,Spathis2011,Quaranta2011,Giazotto2011,Ronzani2013,Ronzani2014,Ligato2017}. 
Figure \ref{fig:vandev}a shows the pseudo-color SEM of a representative V Josephson weak-link along with the four-probes biasing scheme used for the low-temperature characterization \cite{Puglia2020a}.

The device shows a normal-state resistance of about 110 $\Omega$ and switching and retrapping currents of 1.4 mA and 0.35 mA, respectively, at a temperature of 2 K. The evolution of the switching current as a function of the bath temperature, shown in Figure \ref{fig:vandev}b, follows Bardeen's law (red dotted line in Figure \ref{fig:vandev}b). The fit procedure yielded a critical temperature $T_C \simeq 3.62$ K and $I_S^0\simeq2.2$ mA, consistently with the expectation for vanadium-based devices \cite{Spathis2011,Ronzani2013,Ligato2017,Quaranta2011,Ronzani2014,Garcia2009}.

The evolution of the switching current as a function of the gate voltage applied through the gate electrode is shown in Figure~\ref{fig:vandev}c. The curves, acquired at different bath temperatures from 2 K to 3.3 K, show a monotonic suppression of $I_S$, with a complete quenching of the critical current occurring for $ V^C_G \simeq 34 $ V. Also in V devices the widening of the $ I_S $ plateau with temperature was observed \cite{Paolucci2018,DeSimoni2018,DeSimoni2019,Paolucci2019,Bours2020,Rocci2020}.

\begin{figure}[!ht]
\centering
\includegraphics[width=\textwidth]{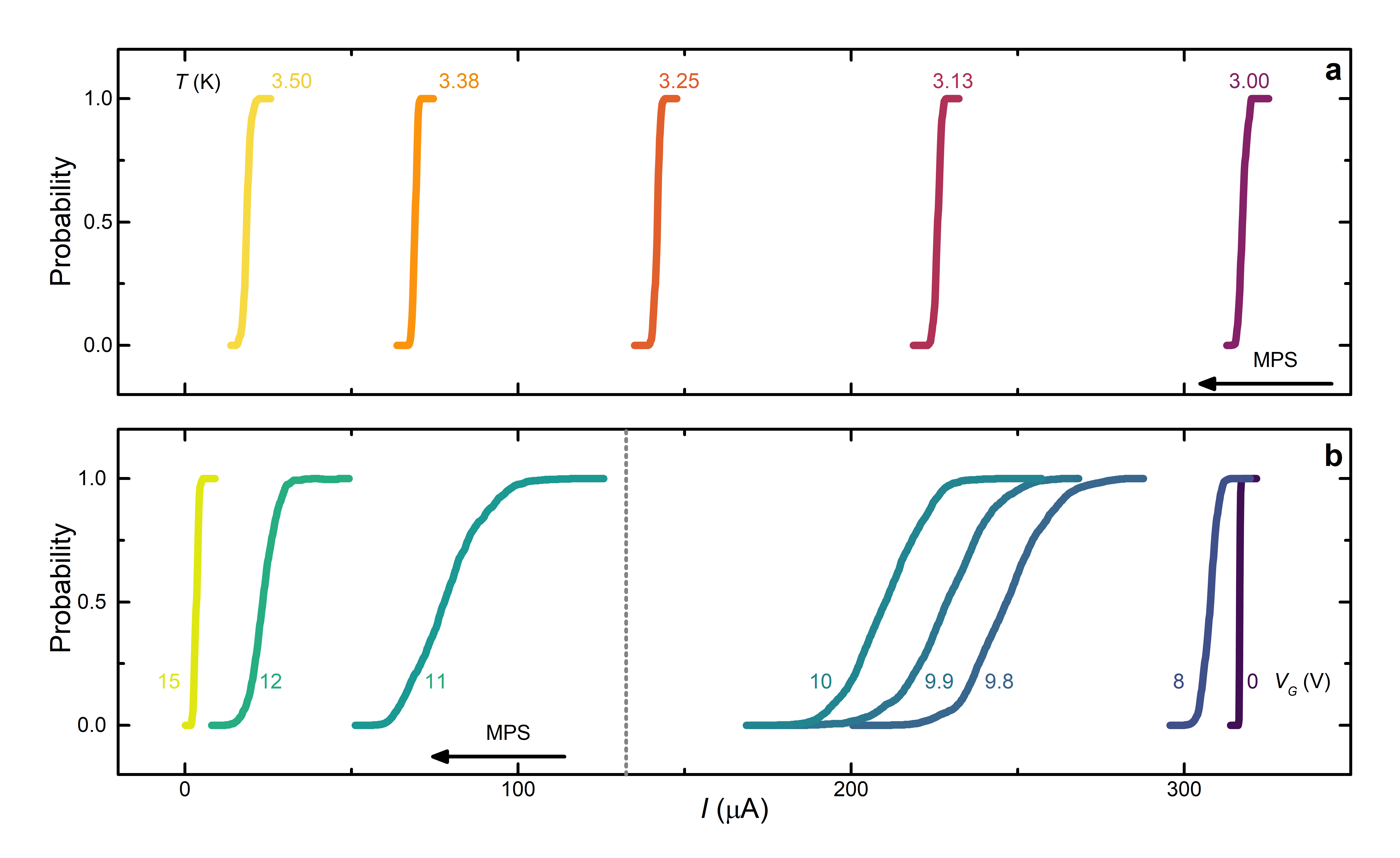}
\caption{a. SCCPDs acquired at several bath temperaturse from 3 K to 3.5 K. The temperature range is compatible with the MPS regime \cite{Bezryadin2012}. b. SCCPDs acquired at several gate voltages from 0 to 15 V at  3.0 K. The curves show starkly different shape due to gate-induced phase slips that widen the zero-to-one probability transition.}
\label{fig:compvan}
\end{figure}

We focus now on the SCCPD behavior as a function of bath temperature and the gate voltage. In particular, we analyze the S-curves with bias current smaller than 330 $\mu$A. Notably, the gate-induced characteristics present a somewhat wider transition than the thermal ones in the same range of switching current. Such a characteristic, already observed in experiments on Ti weak-links, is present in a temperature interval as large as $ T > 3 $ K, i.e., where thermal coupling with the lattice phonons and thermal fluctuations are much larger than in Ti devices \cite{Puglia2020}. This proves that the gating action can dramatically increase the occurrence of phase slips events also in presence of an efficient coupling with the phononic bath near the film critical temperature.

\section{Conclusion}

In this review, we resumed the results of experiments performed on Ti and V gated superconducting mesoscopic weak links. They shed light on the interplay between the conventional gating and the dynamics of phase slips in elemental superconductors Dayem bridges. Firstly, in   Ti-based devices, we showed the impact of the temperature on the shape of the S-curves in a regime where both the electron-phonon coupling and thermal fluctuations were strongly suppressed. At the same time, the gate-driven characteristics present a wider zero-to-one transition for the same value of the critical current compared with the thermal curves. Such SCCPDs comparison is further evidence that the effect of  electrostatic gating cannot be simply interpreted as a  trivial overheating of the junction via current injection or Joule heating.

Secondly, we performed a similar experiment in a  V-based device that presents a critical temperature larger by more than one order of magnitude than the Ti Dayem bridges. The higher critical temperature allowed us to explore a temperature regime where both thermal fluctuations and the electron-phonon coupling are exponentially larger compared to devices operating at sub-kelvin temperatures. Yet, also in this regime, the comparison between thermal- and gate-driven- cumulative switching current probability distributions confirms the results of the previous experiments, highlighting that electrostatic gating affects the dynamics of the phase slips in a starkly different way with respect to temperature.

\section*{Acknowledgments}
We acknowledge the EU’s Horizon 2020 research and innovation program under Grant Agreement No. 800923 (SUPERTED), and the European Research Council under Grant Agreement No. 899315-TERASEC for partial financial support.

%\bibliographystyle{iopart-num}
%\bibliography{references}
%\printbibliography
%\bibliographystyle{unsrt}
%\bibliography{references}

\section*{Bibliography}
\begin{thebibliography}{10}

\bibitem{Lang1970}
N.~D. Lang and W.~Kohn.
\newblock {Theory of Metal Surfaces: Charge Density and Surface Energy}.
\newblock {\em Phys. Rev. B}, 1(12):4555--4568, June 1970.

\bibitem{Larkin1963}
A~I Larkin and A~B Migdal.
\newblock {Theory of superfluid Fermi liquid. Application to the nucleus}.
\newblock Technical Report~5, 1963.

\bibitem{Ummarino2017}
G.~A. Ummarino, E.~Piatti, D.~Daghero, R.~S. Gonnelli, Irina~Yu Sklyadneva,
  E.~V. Chulkov, and R.~Heid.
\newblock {Proximity Eliashberg theory of electrostatic field-effect doping in
  superconducting films}.
\newblock {\em Phys. Rev. B}, 96(6):064509, August 2017.

\bibitem{Shapiro1984}
B.Ya. Shapiro.
\newblock {Surface superconductivity induced by an external electric field}.
\newblock {\em Phys. Lett. A}, 105(7):374--376, October 1984.

\bibitem{Burlachkov1993}
L.~Burlachkov, I.~B. Khalfin, and B.~Ya Shapiro.
\newblock {Increase of the critical current by an external electric field in
  high-temperature superconductors}.
\newblock {\em Phys. Rev. B}, 48(2):1156--1159, July 1993.

\bibitem{Lee1996}
W.D. Lee, J.L. Chen, T.J. Yang, and Bi-Shiou Chiou.
\newblock {Influence of an external electric field on high-Tc
  superconductivity}.
\newblock {\em Phys. C Supercond.}, 261(1-2):167--172, April 1996.

\bibitem{Morawetz2008}
K.~Morawetz, P.~Lipavsk{\'{y}}, J.~Kol{\'{a}}{\v{c}}ek, and E.~H. Brandt.
\newblock {Surface energy and magnetocapacitance of superconductors under
  electric field bias}.
\newblock {\em Phys. Rev. B}, 78(5):054525, August 2008.

\bibitem{Lipavsky2006}
P.~Lipavsk{\'{y}}, K.~Morawetz, J.~Kol{\'{a}}{\v{c}}ek, and T.~J. Yang.
\newblock {Ginzburg-Landau theory of superconducting surfaces under the
  influence of electric fields}.
\newblock {\em Phys. Rev. B}, 73(5):052505, February 2006.

\bibitem{DeSimoni2018}
Giorgio {De Simoni}, Federico Paolucci, Paolo Solinas, Elia Strambini, and
  Francesco Giazotto.
\newblock {Metallic supercurrent field-effect transistor}.
\newblock {\em Nat. Nanotechnol.}, 13(9):802--805, 2018.

\bibitem{Paolucci2018}
Federico Paolucci, Giorgio {De Simoni}, Elia Strambini, Paolo Solinas, and
  Francesco Giazotto.
\newblock {Ultra-Efficient Superconducting Dayem Bridge Field-Effect
  Transistor}.
\newblock {\em Nano Lett.}, 18(7):4195--4199, 2018.

\bibitem{Paolucci2019}
Federico Paolucci, Giorgio {De Simoni}, Paolo Solinas, Elia Strambini, Nadia
  Ligato, Pauli Virtanen, Alessandro Braggio, and Francesco Giazotto.
\newblock {Magnetotransport Experiments on Fully Metallic Superconducting
  Dayem-Bridge Field-Effect Transistors}.
\newblock {\em Phys. Rev. Appl.}, 11(2):024061, February 2019.

\bibitem{DeSimoni2019}
Giorgio {De Simoni}, Federico Paolucci, Claudio Puglia, and Francesco Giazotto.
\newblock {Josephson Field-Effect Transistors Based on All-Metallic Al/Cu/Al
  Proximity Nanojunctions}.
\newblock {\em ACS Nano}, 13(7):7871--7876, July 2019.

\bibitem{Rocci2020}
Mirko Rocci, Giorgio {De Simoni}, Claudio Puglia, Davide Degli~Esposti, Elia
  Strambini, Valentina Zannier, Lucia Sorba, and Francesco Giazotto.
\newblock {Gate-Controlled Suspended Titanium Nanobridge Supercurrent
  Transistor}.
\newblock {\em ACS Nano}, 14(10):12621--12628, October 2020.

\bibitem{Paolucci2019b}
Federico Paolucci, Francesco Vischi, Giorgio {De Simoni}, Claudio Guarcello,
  Paolo Solinas, and Francesco Giazotto.
\newblock {Field-Effect Controllable Metallic Josephson Interferometer}.
\newblock {\em Nano Lett.}, 19(9):6263--6269, September 2019.

\bibitem{Pekker2009}
David Pekker, Nayana Shah, Mitrabhanu Sahu, Alexey Bezryadin, and Paul~M.
  Goldbart.
\newblock {Stochastic dynamics of phase-slip trains and
  superconductive-resistive switching in current-biased nanowires}.
\newblock {\em Phys. Rev. B}, 80(21):214525, December 2009.

\bibitem{Foltyn2015}
Marek Foltyn and MacIej Zgirski.
\newblock {Gambling with Superconducting Fluctuations}.
\newblock {\em Phys. Rev. Appl.}, 4(2):024002, August 2015.

\bibitem{Bezryadin2000}
Alexey Bezryadin, C.~N. Lau, and M.~Tinkham.
\newblock {Quantum suppression of superconductivity in ultrathin nanowires}.
\newblock {\em Nature}, 404(6781):971--974, 2000.

\bibitem{Bezryadin2010}
Alexey Bezryadin and Paul~M. Goldbart.
\newblock {Superconducting Nanowires Fabricated Using Molecular Templates}.
\newblock {\em Adv. Mater.}, 22(10):1111--1121, March 2010.

\bibitem{Bezryadin2012}
Alexey Bezryadin.
\newblock {\em {Superconductivity in Nanowires: Fabrication and Quantum
  Transport}}.
\newblock Wiley-VCH, 2012.

\bibitem{Zgirski2018}
MacIej Zgirski, Marek Foltyn, A.~Savin, K.~Norowski, M.~Meschke, and J.~Pekola.
\newblock {Nanosecond Thermometry with Josephson Junctions}.
\newblock {\em Phys. Rev. Appl.}, 10(4):044068, October 2018.

\bibitem{Virtanen2018}
Pauli Virtanen, Antonio Ronzani, and Francesco Giazotto.
\newblock {Josephson Photodetectors via Temperature-to-Phase Conversion}.
\newblock {\em Phys. Rev. Appl.}, 9(5):54027, 2018.

\bibitem{Solinas2020}
Paolo Solinas, Andrea Amoretti, and Francesco Giazotto.
\newblock {Schwinger effect in a Bardeen-Cooper-Schrieffer superconductor}.
\newblock {\em arXiv}, July 2020.

\bibitem{Mercaldo2020}
Maria~Teresa Mercaldo, Paolo Solinas, Francesco Giazotto, and Mario Cuoco.
\newblock {Electrically Tunable Superconductivity Through Surface Orbital
  Polarization}.
\newblock {\em Phys. Rev. Appl.}, 14(3):034041, September 2020.

\bibitem{Mooij2005}
J.~E. Mooij and C~J P~M Harmans.
\newblock {Phase-slip flux qubits}.
\newblock {\em New J. Phys.}, 7:219--219, October 2005.

\bibitem{Barone1982}
Antonio Barone and Gianfranco Patern{\`{o}}.
\newblock {\em {Physics and Applications of the Josephson Effect}}.
\newblock Wiley, July 1982.

\bibitem{Josephson1962}
B.D. Josephson.
\newblock {Possible new effects in superconductive tunnelling}.
\newblock {\em Phys. Lett.}, 1(7):251--253, July 1962.

\bibitem{Tinkham2004}
Michael Tinkham.
\newblock {\em {Introduction to Superconductivity}}.
\newblock Dover Publications, Mineola, 2004.

\bibitem{Longobardi2011}
Luigi Longobardi, Davide Massarotti, Giacomo Rotoli, Daniela Stornaiuolo,
  Gianpaolo Papari, Akira Kawakami, Giovanni~Piero Pepe, Antonio Barone, and
  Francesco Tafuri.
\newblock {Thermal hopping and retrapping of a Brownian particle in the tilted
  periodic potential of a NbN/MgO/NbN Josephson junction}.
\newblock {\em Phys. Rev. B}, 84(18):184504, November 2011.

\bibitem{Zaikin1997}
Andrei~D. Zaikin, Dmitrii~S. Golubev, Anne van Otterlo, and Gergely~T.
  Zim{\'{a}}nyi.
\newblock {Quantum Phase Slips and Transport in Ultrathin Superconducting
  Wires}.
\newblock {\em Phys. Rev. Lett.}, 78(8):1552--1555, February 1997.

\bibitem{Sahu2009}
Mitrabhanu Sahu, Myung-Ho Bae, Andrey Rogachev, David Pekker, Tzu~Chieh Wei,
  Nayana Shah, Paul~M. Goldbart, and Alexey Bezryadin.
\newblock {Individual topological tunnelling events of a quantum field probed
  through their macroscopic consequences}.
\newblock {\em Nat. Phys.}, 5(7):503--508, 2009.

\bibitem{Ejrnaes2019}
M.~Ejrnaes, D.~Salvoni, L.~Parlato, D.~Massarotti, Roberta Caruso, Francesco
  Tafuri, X.~Y. Yang, L.~X. You, Z.~Wang, G.~P. Pepe, and R.~Cristiano.
\newblock {Superconductor to resistive state switching by multiple fluctuation
  events in NbTiN nanostrips}.
\newblock {\em Sci. Rep.}, 9(1):8053, December 2019.

\bibitem{Kramers1940}
H.A. Kramers.
\newblock {Brownian motion in a field of force and the diffusion model of
  chemical reactions}.
\newblock {\em Physica}, 7(4):284--304, April 1940.

\bibitem{Golubov1999}
A.A. Golubov, K.~Neurohr, Th. Sch{\"{a}}pers, H.~L{\"{u}}th, and M.~Behet.
\newblock {Suppression of Josephson currents in ballistic junctions by an
  injection current}.
\newblock {\em Superlattices Microstruct.}, 25(5-6):1033--1040, May 1999.

\bibitem{Giordano1988}
N.~Giordano.
\newblock {Evidence for Macroscopic Quantum Tunneling in One-Dimensional
  Superconductors}.
\newblock {\em Phys. Rev. Lett.}, 61(18):2137--2140, October 1988.

\bibitem{Kurkijarvi1972}
Juhani Kurkij{\"{a}}rvi.
\newblock {Intrinsic Fluctuations in a Superconducting Ring Closed with a
  Josephson Junction}.
\newblock {\em Phys. Rev. B}, 6(3):832--835, August 1972.

\bibitem{Fulton1971}
T.A. Fulton and R.C. Dynes.
\newblock {Switching to zero voltage in Josephson tunnel junctions}.
\newblock {\em Solid State Commun.}, 9(13):1069--1073, July 1971.

\bibitem{Garg1995}
Anupam Garg.
\newblock {Escape-field distribution for escape from a metastable potential
  well subject to a steadily increasing bias field}.
\newblock {\em Phys. Rev. B}, 51(21):15592--15595, June 1995.

\bibitem{Bardeen1957}
John Bardeen, L.~N. Cooper, and J.~R. Schrieffer.
\newblock {Microscopic Theory of Superconductivity}.
\newblock {\em Phys. Rev.}, 106(1):162--164, April 1957.

\bibitem{Bardeen1962}
John Bardeen.
\newblock {Critical Fields and Currents in Superconductors}.
\newblock {\em Rev. Mod. Phys.}, 34(4):667--681, October 1962.

\bibitem{Bours2020}
Lennart Bours, Maria~Teresa Mercaldo, Mario Cuoco, Elia Strambini, and
  Francesco Giazotto.
\newblock {Unveiling mechanisms of electric field effects on superconductors by
  a magnetic field response}.
\newblock {\em Phys. Rev. Res.}, 2(3):033353, September 2020.

\bibitem{Puglia2020}
Claudio Puglia, Giorgio {De Simoni}, and Francesco Giazotto.
\newblock {Electrostatic Control of Phase Slips in Ti Josephson
  Nanotransistors}.
\newblock {\em Phys. Rev. Appl.}, 13(5):054026, May 2020.

\bibitem{Fulton1974}
T.~A. Fulton and L~N Dunkleberger.
\newblock {Lifetime of the zero-voltage state in Josephson tunnel junctions}.
\newblock {\em Phys. Rev. B}, 9(11):4760--4768, June 1974.

\bibitem{Bae2012}
Myung-Ho Bae, R.~C. Dinsmore, M.~Sahu, and Alexey Bezryadin.
\newblock {Stochastic and deterministic phase slippage in quasi-one-dimensional
  superconducting nanowires exposed to microwaves}.
\newblock {\em New J. Phys.}, 14(4):043014, April 2012.

\bibitem{Blackburn2016}
James~A. Blackburn, Matteo Cirillo, and Niels Gr{\o}nbech-Jensen.
\newblock {A survey of classical and quantum interpretations of experiments on
  Josephson junctions at very low temperatures}.
\newblock {\em Phys. Rep.}, 611:1--33, February 2016.

\bibitem{Garcia2009}
C{\'{e}}sar~Pascual Garc{\'{i}}a and Francesco Giazotto.
\newblock {Josephson current in nanofabricated V/Cu/V mesoscopic junctions}.
\newblock {\em Appl. Phys. Lett.}, 94(13):132508, March 2009.

\bibitem{Spathis2011}
P~Spathis, Subhajit Biswas, Stefano Roddaro, Lucia Sorba, Francesco Giazotto,
  and Fabio Beltram.
\newblock {Hybrid InAs nanowire–vanadium proximity SQUID}.
\newblock {\em Nanotechnology}, 22(10):105201, March 2011.

\bibitem{Quaranta2011}
O.~Quaranta, P.~Spathis, Fabio Beltram, and Francesco Giazotto.
\newblock {Cooling electrons from 1 to 0.4 K with V-based nanorefrigerators}.
\newblock {\em Appl. Phys. Lett.}, 98(3):032501, January 2011.

\bibitem{Giazotto2011}
Francesco Giazotto, Panayotis Spathis, Stefano Roddaro, Subhajit Biswas, Fabio
  Taddei, Michele Governale, and Lucia Sorba.
\newblock {A Josephson quantum electron pump}.
\newblock {\em Nat. Phys.}, 7(11):857--861, November 2011.

\bibitem{Ronzani2013}
Alberto Ronzani, Matthieu Baillergeau, Carles Altimiras, and Francesco
  Giazotto.
\newblock {Micro-superconducting quantum interference devices based on V/Cu/V
  Josephson nanojunctions}.
\newblock {\em Appl. Phys. Lett.}, 103(5):052603, July 2013.

\bibitem{Ronzani2014}
Alberto Ronzani, Carles Altimiras, and Francesco Giazotto.
\newblock {Balanced double-loop mesoscopic interferometer based on Josephson
  proximity nanojunctions}.
\newblock {\em Appl. Phys. Lett.}, 104(3):032601, January 2014.

\bibitem{Ligato2017}
Nadia Ligato, Giampiero Marchegiani, Pauli Virtanen, Elia Strambini, and
  Francesco Giazotto.
\newblock {High operating temperature in V-based superconducting quantum
  interference proximity transistors}.
\newblock {\em Sci. Rep.}, 7(1):8810, December 2017.

\bibitem{Puglia2020a}
Claudio Puglia, Giorgio {De Simoni}, Nadia Ligato, and Francesco Giazotto.
\newblock {Vanadium gate-controlled Josephson half-wave nanorectifier}.
\newblock {\em Appl. Phys. Lett.}, 116(25):252601, June 2020.

\end{thebibliography}

\end{document}